\documentclass[%
 aip,
 amsmath,amssymb,
 reprint,%
]{revtex4-2}
\usepackage{latexsym}
\usepackage{hyperref}
\usepackage{graphicx}%
\usepackage{xcolor}
\usepackage{subfigure} 
\usepackage{dcolumn}%
\usepackage{bm}%
\draft 
\usepackage[utf8]{inputenc}
\usepackage[T1]{fontenc}
\usepackage{mathptmx}
\usepackage{float}
\usepackage{gensymb}
\begin{document}


\title{Molecular Beam Homoepitaxy of N-polar AlN on bulk AlN substrates}

\author{Jashan Singhal}
\email{js3452@cornell.edu}
\affiliation{School of Electrical and Computer Engineering, Cornell University, Ithaca, New York 14853, USA,}

\author{Jimy Encomendero}
\email{js3452@cornell.edu}
\affiliation{School of Electrical and Computer Engineering, Cornell University, Ithaca, New York 14853, USA,}

\author{Yongjin Cho}
\email{js3452@cornell.edu}
\affiliation{School of Electrical and Computer Engineering, Cornell University, Ithaca, New York 14853, USA,}

\author{Len van Deurzen}
\email{js3452@cornell.edu}
\affiliation{School of Applied and Engineering Physics, Cornell University, Ithaca, New York 14853, USA,}

\author{Zexuan Zhang}
\email{js3452@cornell.edu}
\affiliation{School of Electrical and Computer Engineering, Cornell University, Ithaca, New York 14853, USA,}

\author{Kazuki Nomoto}%
\affiliation{School of Electrical and Computer Engineering, Cornell University, Ithaca, New York 14853, USA,}

\author{Masato Toita}%
\affiliation{Advanced Devices Technology Center, Asahi Kasei Corporation, Hibiya Mitsui Tower, 1-1-2 Yurakucho, Chiyodaku, Tokyo 100-8440, Japan,}

\author{Huili Grace Xing}%
\affiliation{School of Electrical and Computer Engineering, Cornell University, Ithaca, New York 14853, USA,}
\affiliation{Department of Materials Science and Engineering, Cornell University, Ithaca, New York 14853, USA}
\affiliation{Kavli Institute at Cornell for Nanoscale Science, Cornell University, Ithaca, New York 14853, USA}

\author{Debdeep Jena}%
\affiliation{School of Electrical and Computer Engineering, Cornell University, Ithaca, New York 14853, USA,}
\affiliation{Department of Materials Science and Engineering, Cornell University, Ithaca, New York 14853, USA}
\affiliation{Kavli Institute at Cornell for Nanoscale Science, Cornell University, Ithaca, New York 14853, USA}





\date{\today}

\begin{abstract}
N-polar AlN epilayers were grown on the N-face of single crystal bulk AlN substrates by plasma assisted molecular beam epitaxy (PA-MBE). A combination of \emph{in situ} thermal deoxidation and Al-assisted thermal desorption at high temperature helped in removing native surface oxides and impurities  from the N-polar surface of the substrate enabling successful homoepitaxy. Subsequent epitaxial growth of AlN layer on the \emph{in situ} cleaned substrates, grown in sufficiently high Al droplet regime, exhibited smooth surface morphologies with clean and wide atomic steps. KOH etch studies confirmed the N-polarity of the homoepitaxial films. Secondary ion mass spectrometry profiles show Si and H impurity concentrations below the noise levels, whereas O and C impurities concentrations of  $\sim 8 \times 10^{17}$ atoms/cm$^3$ and $\sim 2\times 10^{17}$ atoms/cm$^3$ are observed respectively. Though the structural defect densities are low, they interestingly appear as inversion domains of different dimensionalities.
\end{abstract}

\maketitle 


Due to the absence of inversion symmetry, the metal-polar ($0001$) and N-polar ($000\bar{1}$) surfaces of wurtzite III-nitrides are not equivalent. As a result, epitaxial films grown on the metal-face and the N-face exhibit different surface morphologies, optical properties and growth kinetics. \cite{monroy2004growth, held1999structure} 
Growth of N-polar nitrides is particularly challenging compared to their metal-polar counterparts. The first challenge is that the adsorption efficiency of impurities such as oxygen is higher in N-polar surfaces compared to the metal polar ones.\cite{zywietz1999adsorption,takeuchi2007improvement,ye2008asymmetry,miao2010effects} Second, a low surface diffusivity \cite{zywietz1998adatom, scheffler2003adatom} on the N-polar face is thought to inhibit step-flow-growth mode. 

The growth of smooth, high quality N-polar nitride films is technologically relevant for the development of high electron mobility transistors (HEMTs) built on the wide-bandgap III-nitride materials platform for high power mm-wave electronics. In fact, current state-of-the-art performance has been achieved on the N-polar GaN platform in N-polar GaN/AlGaN HEMTs with output powers above 8 W/mm at up to 94 GHz.\cite{romanczyk2020w} While the transistors based on N-polar GaN platform have proven to be a useful device technology with their excellent performance in W-band (75-110 GHz) in the past decade\cite{wong2013n,keller2014recent}, the replacement of N-polar GaN with N-polar AlN as the buffer layer can enable further advancements. There has been significant interest recently in the development of RF electronics on the AlN platform \cite{hickman2021next} due to the ultrawide and direct bandgap (6 eV), high thermal conductivity ($\sim 340$ W/mK) and high piezoelectricity of AlN.  N-polar AlN can enable the formation of two-dimensional electron gas (2DEG) with AlN back-barriers, with the promise of enhanced thermal management, reduced buffer leakage and short channel effects.\cite{wong2008n, lemettinen2018transistor, lemettinen2019nitrogen} High conductivity 2DEG channels with high density 2DEGs  induced at the heterointerface with AlN can then be expected.\cite{wong2008n, lemettinen2018transistor, smorchkova2001aln}

Homoepitaxial growth of high quality N-polar AlN is an important first step for N-polar RF electronics on the AlN platform. Growth of N-polar wurtzite AlN  on foreign substrates such as Si, SiC and sapphire
by metal-organic vapor phase epitaxy (MOVPE), sputtering
and molecular beam epitaxy (MBE) have been reported.\cite{keller2006effect,takeuchi2007and,dasgupta2009growth,okumura2012growth,ledyaev2014Sin,lemettinen2018movpe,isono2020growthAlN,shojiki2021reduction} Recently, MBE homoepitaxy of N-polar AlN grown on AlN templates on sapphire has also been demonstrated. \cite{Zexuan2022} A limited
number of reports exist on the growth of N-polar AlN on native AlN substrates.\cite{isono2020growthAlN} This is because single crystal AlN substrates with epi-ready N-face have not been readily available. Growth on substrates where lattice and thermal mismatch is present leads to a high density of threading dislocations and other defects. These defects act as scattering centers and traps for carriers and hamper the performance of devices. There is a strong desire to study the growth of N-polar AlN on bulk AlN substrates to eliminate the above problems. Moreover, a homoepitaxial platform for N-polar AlN could offer new possibilities in deep-ultraviolet (UV) laser diodes and light-emitting diodes \cite{Zhang_2019,Len2022}. Besides suppressing optical losses due to the reduction in dislocation and point defect densities by growth on single-crystal AlN \cite{Tanaka_2020}, the N-polarity can improve the light output in deep-UV emitters by enhancing the injection efficiency \cite{Henryk2019}.    While the successful molecular beam homoepitaxy of \emph{metal}-polar AlN on single-crystal AlN substrates has been recently demonstrated \cite{YongjinAlpolish,KevinAlpolish}, the study of the homoepitaxy of N-polar AlN on bulk substrates has been lacking.

In this work, the MBE homoepitaxy of N-polar
AlN films on bulk N-polar AlN substrates is reported. Taking inspiration from the \emph{in situ} cleaning techniques of thermal annealing and Al assisted cleaning developed for Al polar AlN \cite{YongjinAlpolish, KevinAlpolish}, an \emph{in situ} cleaning method for the surface of bulk N-polar AlN is developed, before performing epitaxy. Careful monitoring of \emph{in situ} reflection high-energy electron diffraction (RHEED) helps to repeatably obtain clean N-face AlN surfaces free of contaminants and enables a meaningful study of homoepitaxy of N-polar AlN.

Subsequently, AlN homoepitaxy is performed by PA-MBE. The surface morphology after epitaxy is found to be strongly dependent on the Al droplet coverage during growth where the condition with the highest Al droplet accumulation is found to produce parallel atomic steps.  On the other hand, at low Al droplet coverage, it is observed that the surface suffers from pits and spiral hillocks. This is likely because Al droplets facilitate adatom diffusion on the N-face.   KOH etching study is used to confirm the N-polarity of the homoepitaxial films while also revealing a low density of Al-polar inversion domains whose properties are discussed in detail.  Finally, secondary ion mass spectrometry (SIMS) analysis of the smooth homoepitaxial N-polar AlN films indicates low chemical impurity densities for hydrogen and silicon near the detection limit, whereas oxygen and carbon show higher densities of incorporation.
\begin{figure}[!htbp]
\includegraphics[width = 0.4\textwidth]{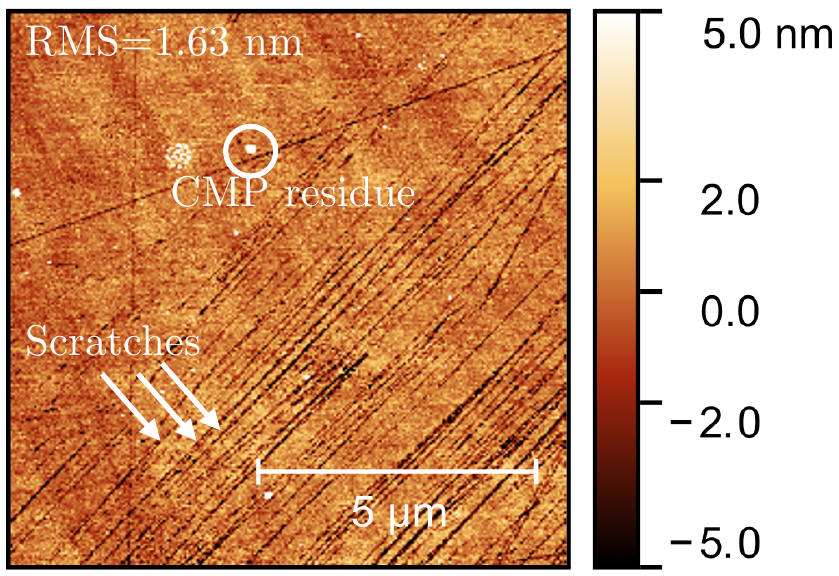}
\caption{\label{before_growth} AFM micrograph of the surface of the N-face of an as received bulk AlN substrate. The surface has particulate residue and scratches from the CMP process. }
\end{figure}

2-inch diameter N-polar AlN ($000\bar{1}$) bulk substrates prepared by Asahi Kasehi were used in this study. The dislocation densities of these substrates were less than $10^4$ cm$^{-2}$ signifying high structural quality. The substrates have surface defects arising from chemical-mechanical polishing (CMP) process used on the N-face of AlN to prepare it for epitaxy. Figure \ref{before_growth} shows an AFM image of the N-face of the single crystal AlN substrate prior to cleaning and epitaxy. The surface morphology is largely smooth with a root-mean square (rms) roughness between 1-2 nm over a 10 $\mu$m $\times$ 10 $\mu$m area.  But  scratches and particulate debris from the CMP process covering the surface are observed, as indicated in Fig. \ref{before_growth}. The wafers were diced into 7x7 mm$^2$ pieces and cleaned \emph{ex-situ} in acetone, isopropyl alcohol and deionized water. They were then mounted on a faceplate and transferred into the load-lock chamber of a Veeco GEN10
plasma-MBE system and degassed at $200 \degree$C for 8 h. No other \emph{ex situ} cleaning procedures (e.g. using strong acids such as Piranha, H$_2$SO$_4$ etc.) were used here since it was found that such treatments damage the surface of N-face AlN. This is in stark contrast to the \emph{ex situ} cleaning of the Al-face of bulk AlN substrates where acid treatment  reveals the atomic steps on the surface instead. \cite{KevinAlpolish}

\begin{figure}[!htbp]
\includegraphics[width = 0.5\textwidth]{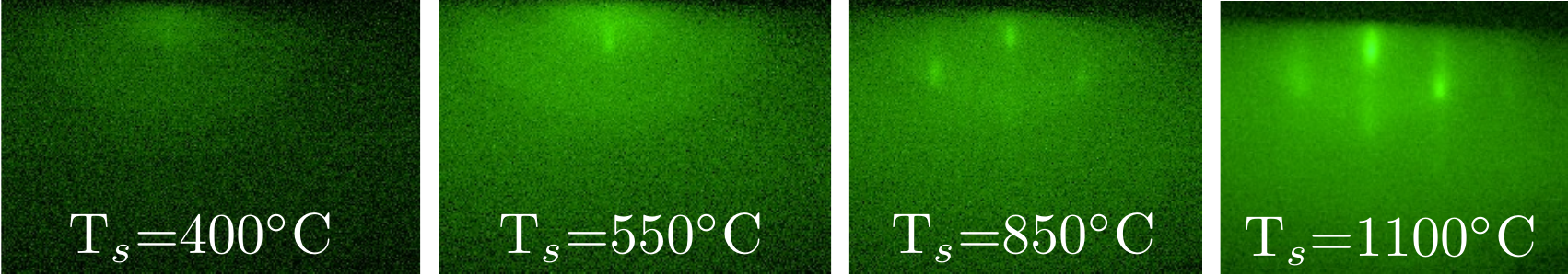}
\caption{\label{thermal_cleaning} Evolution of the RHEED patterns of the AlN substrate taken along $\langle 11\bar{2}0\rangle$ azimuth during thermal deoxidation as the substrate temperature is progressively increased from 400$\degree$C to 1100$\degree$C.   }
\end{figure}

\emph{In situ} cleaning of the substrates was performed in the MBE growth chamber using a combination of thermal deoxidation and Al assisted thermal desorption at a substrate temperature of $\approx$ 1100$\degree$C. Evolution of the surface during the \emph{in situ} cleaning of the AlN substrate, as well as during and after the homoepitaxy of AlN were monitored by RHEED. Firstly, changes in the AlN substrate surface during thermal deoxidation process were studied. The substrates were heated up to a thermocouple temperature of 1100$\degree$C. As the temperature of the substrate was increased from  400$\degree$C to 1100$\degree$C, the RHEED pattern of the sample evolved from dim and diffuse to bright and spotty signifying that the thermally volatile oxides and hydroxides are removed from the surface, as shown in Figure \ref{thermal_cleaning} . The surface becomes less amorphous as a result of this thermal deoxidation process,  but further \emph{in situ} chemical cleaning was found to be necessary to remove thermally non-volatile oxides and impurities.

\begin{figure*}[!htbp]
\includegraphics[width = 1\textwidth]{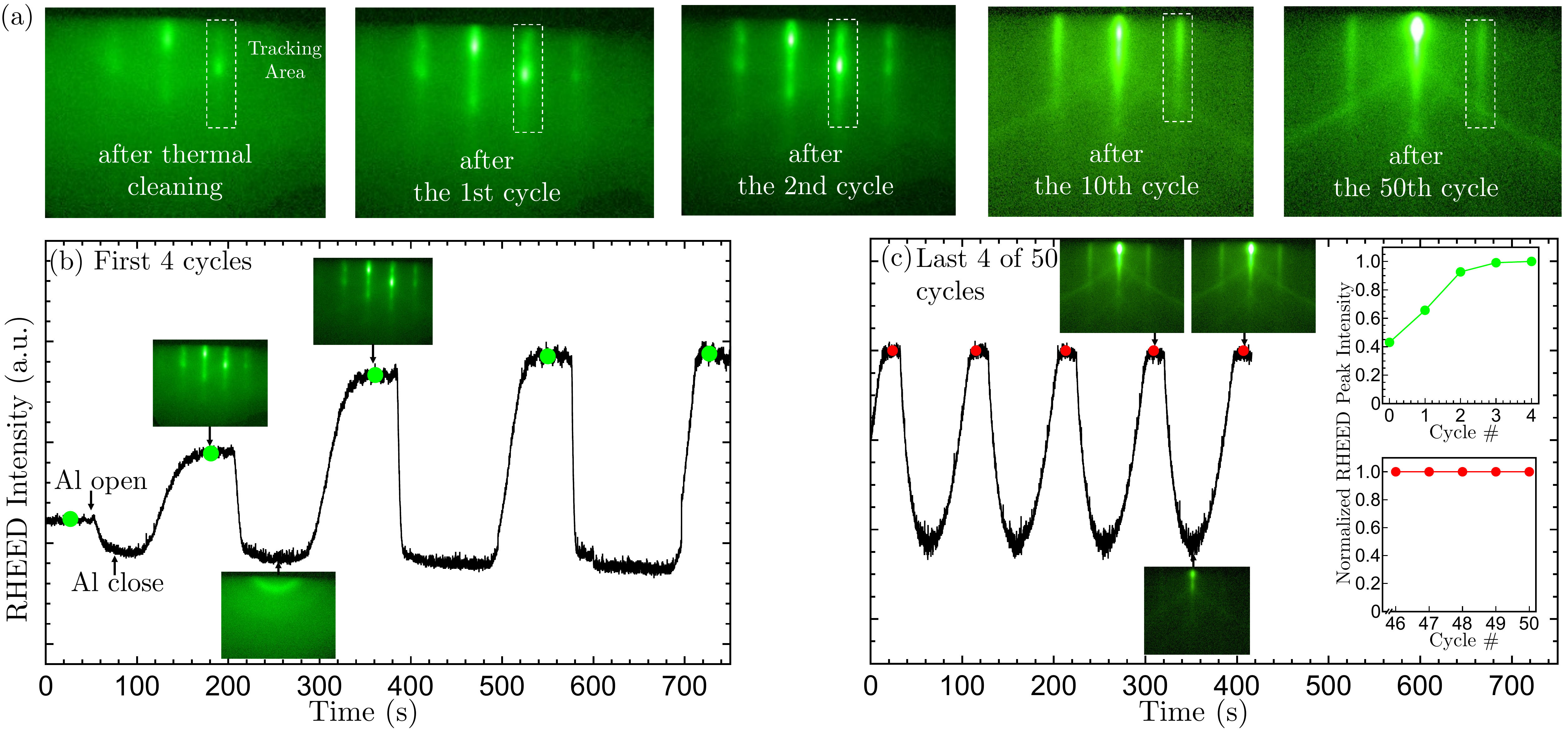}
\caption{\label{Al_polishing}(a) RHEED patterns of the AlN substrate at various stages during the 50 Al polishing cycles. RHEED intensity vs time of the boxed area in (a) is shown in (b) for the first four cycles and (c) for the last four cycles. Inset RHEED images in (b) and (c) show the diffraction patterns during the 2nd and 50th Al polishing cycles respectively. The inset in (c) shows the evolution of normalized RHEED peak intensity after each cycle as a function of the cycle number. All the RHEED patterns were taken along the $\langle 11\bar{2}0\rangle$ azimuth. }
\end{figure*}

While holding the substrate at a thermocouple temperature of 1100$\degree$C, \emph{in situ} Al assisted surface cleaning was then performed to prepare for epitaxy. This process involved exposing the surface to an aluminum metal flux $\phi_{Al}=15.9$ nm/min for 30s, and then holding the high temperature of the substrate long enough for the deposited Al to desorb; this adsorbtion and desorption, which \citeauthor{YongjinAlpolish} developed for the cleaning of Al polar bulk AlN substrates \cite{YongjinAlpolish} and \citeauthor{KevinAlpolish} called Al polishing \cite{KevinAlpolish}, was repeated for $\sim 50$ cycles. Though the procedure of Al polishing of N-polar AlN is similar to the Al-polar single crystal AlN, a few  characteristic differences are observed in the nature of RHEED evolution during the course of Al assisted cleaning.

Figure \ref{Al_polishing} shows the evolution of the RHEED pattern and  the variation of RHEED intensity during the Al polishing cycles. Fig. \ref{Al_polishing}(a) depicts the gradual transformation of the diffraction pattern from spotty diffuse  after thermal cleaning, to bright and streaky  after 50 cycles of Al polishing.
Fig. \ref{Al_polishing}(b) and Fig. \ref{Al_polishing}(c)   track the temporal evolution of the RHEED intensity of the boxed
area in Fig. \ref{Al_polishing}(a), during the first four and last four cycles of Al polishing, respectively. In each cycle, the intensity first drops when Al is deposited. Subsequently, when Al is desorbed, the intensity increases and eventually saturates to the peak value of the cycle. This is similar to the observations made for Al-polar AlN. \cite{YongjinAlpolish,KevinAlpolish} The first three cycles are distinct
as the corresponding peak/saturation intensity progressively increases over them, after which the saturation intensity remains unchanged for the remaining cycles. This gradual increase in the saturation intensity over the initial few cycles can be explained by the removal of amorphous contaminants such as oxides and hydroxides of Al and various other impurity alloys of Al,O and N from the surface of the substrate during Al polishing, thereby improving the crystallinity of the surface. 

The RHEED images in Fig.\ \ref{Al_polishing}(b) also show how the Al metal reacts with the contaminated surface of the substrate during the 2nd cycle of Al polishing. Starting with a spotty and diffuse RHEED pattern before the Al adsorption, the RHEED pattern exhibits diffuse Debye rings when Al reacts with the surface. These rings indicate the formation of a polycrystalline complex which is removed from the surface as Al is desorbed, revealing a visibly brighter and streakier RHEED pattern, implying a smoother and cleaner surface. This process of Al assisted cleaning is repeated for 50 cycles until  no further change is observed in the evolution of RHEED intensity as a function of time as seen in the last four cycles of Al polishing in Fig.\ \ref{Al_polishing}(c). It is noteworthy that when the surface of AlN is sufficiently clean, the Al adsorption and desorption behaves differently compared to when the surface is contaminated. The RHEED images in Fig.\ \ref{Al_polishing}(c) show how the cleaned AlN surface behaves during Al adsorption and desorption in the final (50th) cycle. The RHEED pattern is initially bright and streaky. Upon Al adsorption the RHEED pattern becomes dark due to Al adlayer formation and Al droplet accumulation without forming any polycrystalline complex. The streaky pattern is recovered upon Al desorption. This change in the behavior during Al adsorption accurately signals when the surface is free of contamination, along with the RHEED pattern changing from spotty-diffuse  before Al polishing to bright-streaky after 50 cycles as shown in Fig.\ \ref{Al_polishing}(a).
\begin{table}
\caption{\label{tab:conditions}Growth conditions with nitrogen plasma power at 400 W and 1.85 sccm flow rate}
\begin{ruledtabular}
\begin{tabular}{cccc}
Condition & Substrate Temperature ($\degree$C) & $\phi_{Al}$ (nm/min) & Al/N Ratio\\ \hline
A & $1055$  & $ 12.7$ & 1.84 \\
B & $1055$  & $19.6$ & 2.84 \\
C & $1000$  & $19.6$ &  2.84\\
\end{tabular}
\end{ruledtabular}
\end{table}

\begin{figure*}[!htbp]
\includegraphics[width = 1\textwidth]{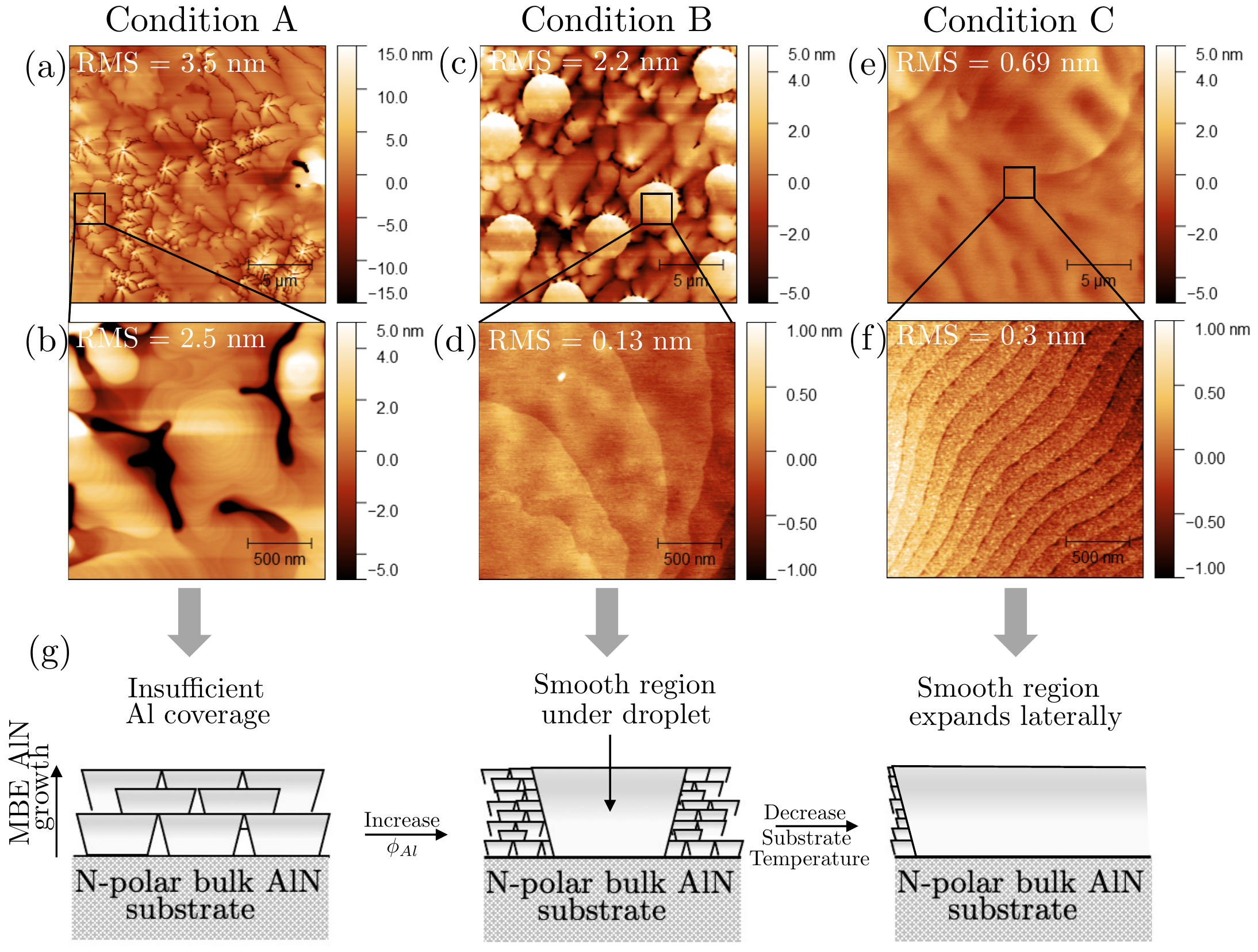}
\caption{\label{homo_AFM} AFM micrographs of AlN grown
on N-polar AlN bulk substrates under various metal-rich growth conditions (see Table \ref{tab:conditions}). AFM images in (a), (c) and (e) show $20\times20 $ $\mu$m$^2$ area scans and (b), (d) and (f) show $2\times2 $ $\mu$m$^2$ area scans of 500 nm AlN grown under conditions \textbf{A}, \textbf{B} and \textbf{C} respectively. (g) Schematic of MBE growth of N-polar AlN under various growth conditions.  }
\end{figure*}

 AlN homoepitaxy was performed directly after the \emph{in situ} cleaning. Surface morphologies of AlN grown in various metal-rich growth conditions (see Table \ref{tab:conditions}) were explored for  pit-free and smooth homoepitaxy. The nitrogen
plasma power was held at 400 W with 1.85 sccm flow rate, corresponding
to a growth rate of 6.9 nm/min ($=\phi_N$) where $\phi_N$ is the N flux. Al metal was provided from a standard effusion cell. AlN films of thicknesses $\sim$500 nm were grown in three conditions shown in Table \ref{tab:conditions}. The epitaxial process was started by
first opening the Al shutter for 20 s to wet the surface. Then Al and N fluxes were provided simultaneously during the epitaxial growth. To end the growth, the plasma is turned off, both Al and nitrogen
shutters are closed, and the substrate is heated to $50\degree$C higher than
the growth temperature to desorb excess Al. 
\begin{figure*}[!htbp]
\includegraphics[width = 1\textwidth]{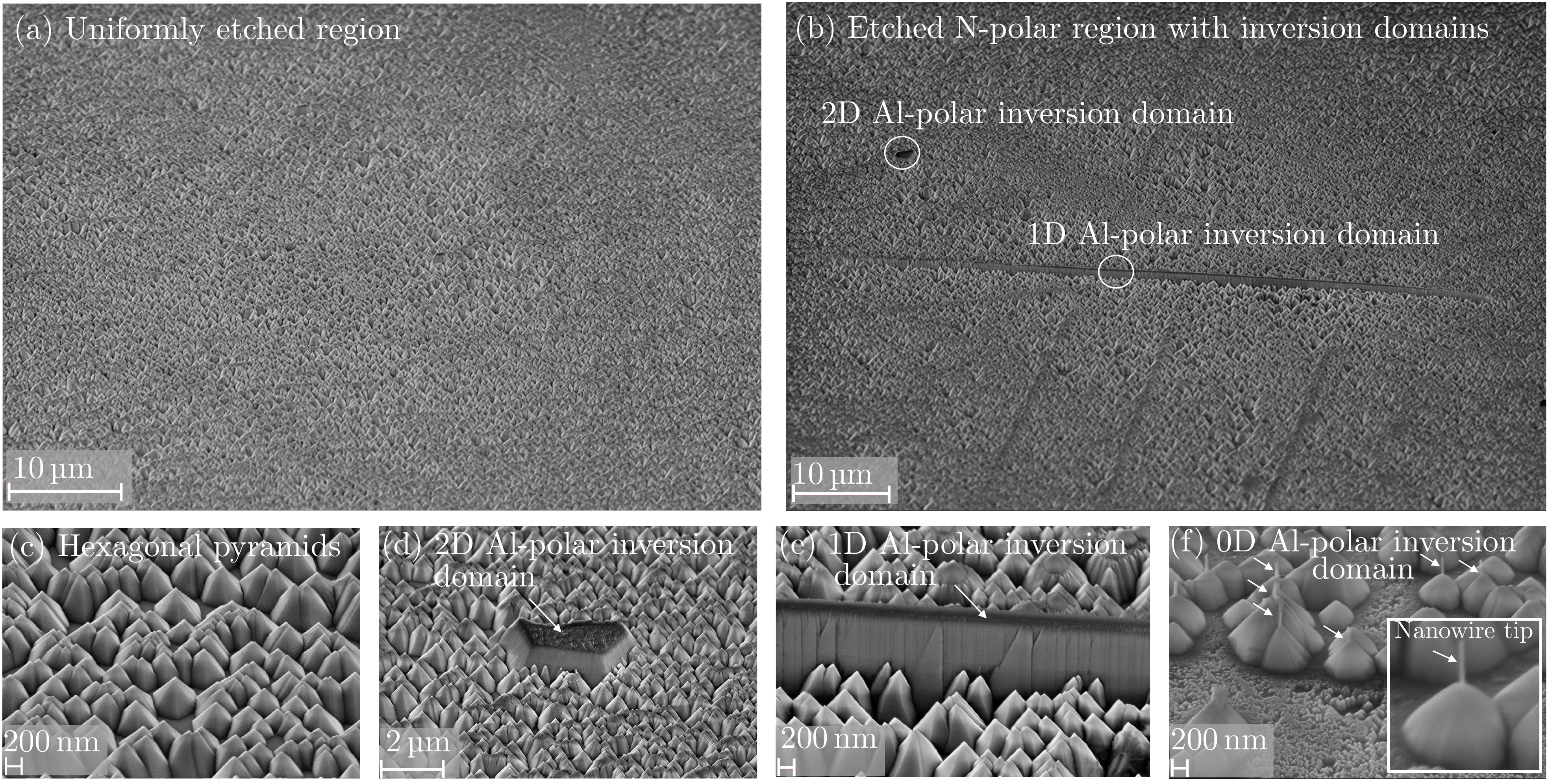}
\caption{\label{KOH_etch} Tilted aerial SEM images of (a) uniformly etched N-polar AlN in KOH, (b) etched region with inversion domains, (c) hexagonal pyramids in the uniformly etched region, (d) 2D Al-polar inversion domain, (e) 1D Al-polar inversion domain and (f) 0D Al-polar inversion domain. Inset of (f) shows a zoomed in image of an inversion nanowire tip.   }
\end{figure*}
It was found that the surface morphology of homoepitaxial N-polar AlN is strongly dependent on the amount of excess Al on the surface during growth. Excess Al droplets can be controlled by the $\phi_{Al}$ impinging on the surface, as well as by controlling the Al droplet desorption rate, which is a function of the substrate temperature. Condition \textbf{A} in Table \ref{tab:conditions} should lead to the lowest droplet accumulation because it is at the highest investigated substrate temperature of $1055\degree$ C and the lowest impinging Al/N ratio. Due to the insufficient Al coverage, a large number of pits were observed in the surface morphology of this sample, as well as spiral hillocks [Figs. \ref{homo_AFM}(a) and \ref{homo_AFM}(b)], implying the formation of dislocations.
The sample grown in condition \textbf{B} at the same substrate temperature of $1055\degree $ C as condition \textbf{A}, but at a  $\phi_{Al}$ that is almost 1.5 times higher that of  condition \textbf{A},  reveals two distinct
regions in the surface morphology. Smooth circular areas with clear atomic steps, and rough regions characterized by surface pits and spiral hillocks [Figs. \ref{homo_AFM}(c) and \ref{homo_AFM}(d)] are observed. The size and shape of the smooth regions indicate that it is possible that the diffusion of the adatoms is improved under the Al droplets, resulting in smooth morphology, indicating that the regions of AlN growth under Al droplets show two-dimensional (2D) growth mode. While keeping the $\phi_{Al}$ same as in condition \textbf{B}, the substrate temperature is reduced to $1000\degree $ C in condition \textbf{C} to reduce the Al desorption rate and to increase Al accumulation. It is observed that this sample shows a smooth surface with clear atomic steps over a large surface area [Figs. \ref{homo_AFM}(e) and \ref{homo_AFM}(f)]. The density of the smooth circular regions that were observed in condition \textbf{B} increased under condition \textbf{C} because of larger Al accumulation. The smooth regions coalesced to enable smooth homoepitaxial growth with clean and wide
atomic terraces on the surface and rms roughness $<0.7$ nm over a 20 $\mu$m $\times$ 20 $\mu$m area. This observation is in sharp contrast with the MBE growth of typical group III-nitrides for which the presence/absence of the metallic droplets do not influence the growth modes as long as a complete metallic adlayer is formed on the surface.\cite{AlNGrowthDiag2003,Koblmuller_2005,Kobmuller2007} For the samples in this study, therefore, it is likely that a relatively high density of surface impurities, which presumably segregate from the substrate surface, hinder the formation of Al adlayer thus requiring Al droplets for smooth surface morphology. A similar effect has been reported in MBE-grown Mg-doped GaN where Mg affects the growth kinetics by making Ga-adlayer unstable. \cite{Monroy2004}

Based on the surface morphologies observed, Fig. \ref{homo_AFM}(g) schematically shows the probable epitaxial evolution of the MBE growth of N-polar AlN under each of the conditions. Further transmission electron microscopy studies would help to understand the nucleation interface as well as the various growth modes better. Thus, the growth at substrate temperature of $\sim 1000\degree$ C, with $\phi_{Al} \sim 19.6$ nm/min, $\phi_{N} \sim 6.9$ nm/min and $\phi_{Al}/\phi_{N} \sim 2.84$ on the cleaned N-polar AlN surface is found to lead to smooth homoepitaxy.

Next, we confirm that the epitaxial layers retain the N-polarity of the underlying single-crystal AlN substrate. Wet chemical etching by KOH was used to determine the lattice polarity and its spatial homogeneity in the as grown AlN epitaxial layers. KOH etching is sensitive to the polarity of (Al,Ga)N. The etching rate and morphological features post etching depend strongly on the chemical nature of the surface.\cite{li2001selective, guo2013comparative, guo2015koh} The
N-polar surface reacts and etches in KOH, leaving hexagonal pyramids bounded by more chemically stable $\{1\bar{1}0\bar{1}\}$ crystallographic planes. The metal polar surface remains fairly inert to KOH and only shows defect-selective etch behavior with hexagonal pits generated around dislocations after etching.\cite{lu2008microstructuremetalpolarKOH} Fig. \ref{KOH_etch} shows 45$\degree$ tilted aerial scanning electron microscopy (SEM) scans of
a 500 nm epitaxial layer grown under condition \textbf{C} on bulk N-polar AlN, after etching in 50 wt\% KOH aqeuous solution at 80$\degree$C for 10 seconds. The surface is vigorously
etched and the etch is nearly uniform in most regions, as seen in Fig.\ref{KOH_etch} (a). It forms hexagonal pyramids of heights ranging from 400-600 nm as seen in  Fig.\ref{KOH_etch} (c). This indicates that nearly the entire epitaxial film maintained the N-polarity of the substrate. 

The etching process also revealed a low density of inversion domains as seen in Fig.\ref{KOH_etch} (b) which are regions that were not etched by KOH. Three types of inversion domains were identified in the sample. They are classified on the basis of their shape and size, as 2D, 1D and 0D Al-polar inversion domains as shown in Figs.\ref{KOH_etch} (d)- \ref{KOH_etch}(f). It is difficult to estimate the density of 2D and 1D Al-polar inversion domains as they are quite sporadic and sparsely spaced in the field of view of the SEM but their combined density is $\sim 1\times10^4$/cm$^2$. 0D Al-polar inversion domains are less common than 2D and 1D ones but wherever they are revealed after etching, they appear in clusters as seen in Fig.\ref{KOH_etch} (f).

In spite of the scarce occurrence of these inversion domains, their properties and possible origins are briefly discussed. They are believed to have formed right at the nucleation stage of AlN on the substrate. They propagate through the entire thickness of the 
epitaxial film during growth, growing in the Al-polar orientation as opposed to the N-polarity of the majority of the film. The formation of these inversion domains is linked to improper substrate surface preparation before epitaxy. A 2D Al-polar inversion domain as shown in Fig.\ref{KOH_etch} (d), has a plateau like feature, approximately 1-2 $\mu$m in lateral dimensions and a height of approximately 500 nm (the thickness of the epi-layer). 1D Al-polar inversion domains as seen in Fig.\ref{KOH_etch} (e) are several tens of microns long  wall-like features. They most likely are the remains of scratches after the CMP process (Fig. \ref{before_growth}) of the as-received substrates. Fig.\ref{KOH_etch} (f) shows 0D Al-polar inversion domains indicated by arrows which are Al-polar nanowire tips protruding from the apex of hexagonal N-polar pyramids. They are similar to the ones reported in previous works on N-polar AlN films on sapphire\cite{InvDomainUCB,InversionDomain1}.  \citeauthor{InvDomainUCB} \cite{InvDomainUCB} attributed the occurrence of such inversion domains in MOCVD N-polar AlN films grown on sapphire substrates to tiny
Al-polar inversion domains nucleating at the substrate, and growing faster than the surrounding N-polar matrix that they are embedded in. It is likely that surface impurities such as oxygen, and particulate residues from CMP on the substrate surface (Fig. \ref{before_growth}) contribute to the polarity inversion by acting as nucleating sites for 2D and 0D inversion domains. It has been demonstrated that oxygen plays a critical role in switching N-to-Al-polarity in AlN grown on sapphire, where the formation of  aluminum-oxynitride (Al$_x$O$_y$N$_z$) acts as the inversion domain  boundary.\cite{stolyarchuk2017impactIDB,stolyarchuk2018intentionalIDB} Even though the density of inversion domains present in our film is quite low, their presence can have negative impact on the electronic and photonic properties of devices made on the N-polar AlN platform. Therefore, continuous improvements in substrate preparation, cleaning and growth recipes are necessary to suppress their formation.

\begin{figure}[!htbp]
\includegraphics[width = 0.5\textwidth]{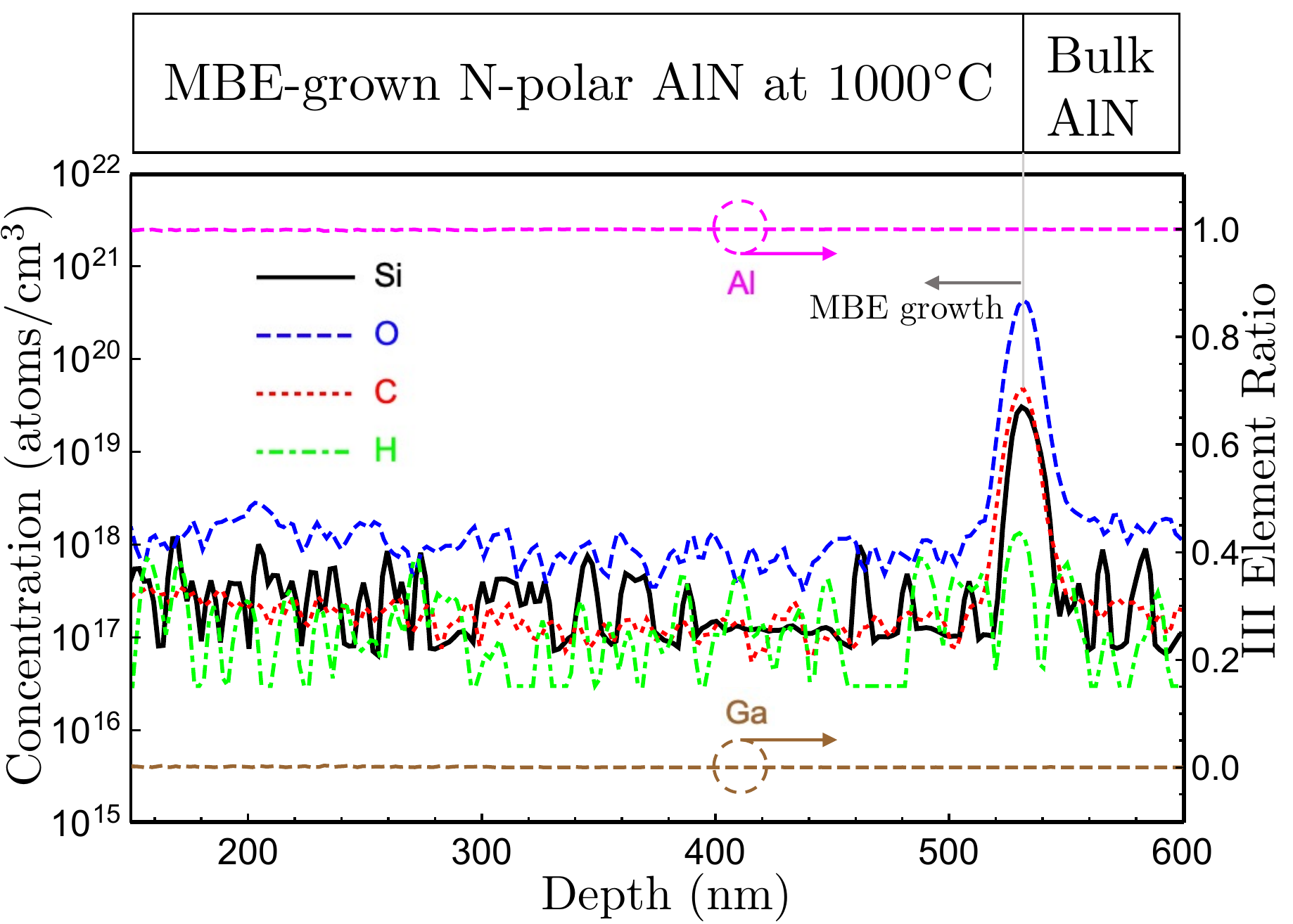}
\caption{\label{SIMS}Secondary ion mass spectrometry profile of silicon, oxygen, hydrogen, and carbon in the N-polar AlN epitaxial films grown by PA-MBE. The peak concentrations for  O, C and Si are $\sim 4.5\times 10^{20}$ atoms/cm$^3$, $\sim 4.8\times 10^{19}$ atoms/cm$^3$ and  $\sim 3.2\times 10^{19}$ atoms/cm$^3$ respectively. The III element ratio is specified
on the right axis. The Ga content is less than 1\% throughout the film, suggesting that there is no Ga contamination in the AlN. }
\end{figure}

A sample with 500 nm of N-polar AlN grown in conditions similar to condition \textbf{C} (See Table \ref{tab:conditions}) was investigated by SIMS. Figure \ref{SIMS} shows that the Si and H impurities are near, or below the detection limit throughout the 500 nm epitaxial layer away from the nucleation interface. O and C impurities show higher concentrations of  $\sim 8 \times 10^{17}$ atoms/cm$^3$ and $\sim 2\times 10^{17}$ atoms/cm$^3$ respectively (the detection limits are $2\times 10^{17}$ atoms/cm$^3$ for Si and O, $1$-$2\times 10^{17}$ atoms/cm$^3$ for H, and $6$-$7\times 10^{16}$ atoms/cm$^3$ for C). Previously reported values for O and C impurity incorporation levels for epitaxial Al-polar AlN films were about $1$-$2\times 10^{17}$ atoms/cm$^3$ and $4$-$5\times 10^{16}$ atoms/cm$^3$ respectively,  grown on the Al-face of single crystal AlN from the same substrate provider.\cite{KevinAlpolish}Similar observations of higher O and C impurity incorporation in N-polar nitride films compared to metal-polar ones have been reported for both MOCVD AlN on sapphire\cite{takeuchi2007and}, and also for MOCVD GaN\cite{GaNimpuritypol}  previously. Previous experimental and theoretical studies have indicated that oxygen atoms favorably adsorb on N-polar GaN and AlN surfaces than on metal polar ones because the adsorption energy at high oxygen coverage is significantly lower for the N-polar surfaces compared to their metal-polar counterpart.\cite{zywietz1999adsorption,takeuchi2007improvement,ye2008asymmetry,miao2010effects} Carbon impurities in the epitaxial N-polar nitride films cause the generation of surface pits in films grown on N-polar GaN subtrate.\cite{WURM2020113763}

 It is evident from Figure \ref{SIMS} that  there  is  a  buildup  of  impurities  at  the  growth interface. In  addition  to  the residual impurities  left  over  from  polishing  and  processing  performed before epitaxy, impurities may also accumulate on the surface during exposure to atmosphere. Peaks for O, C and Si were observed at the nucleation interface with concentrations of $\sim 4.5\times 10^{20}$ atoms/cm$^3$, $\sim 4.8\times 10^{19}$ atoms/cm$^3$ and  $\sim 3.2\times 10^{19}$ atoms/cm$^3$ respectively. It was observed that thermal annealing of Al-polar AlN single crystal substrate prior to homoepitaxy at $\sim 1450\degree$C can nearly remove all carbon from the substrate surface \cite{YongjinAlpolish}. Thus, such high temperature annealing can be possibly employed to suppress C incorporation in N-polar AlN in future.

In conclusion, these results indicate that \emph{in situ} thermal deoxidation and Al assisted cleaning  enables the smooth epitaxial growth of N-polar AlN on single crystal AlN substrates. It is shown that the surface morphology of homoepitaxial N-polar AlN is sensitive to the amount of Al droplet accumulation during the metal-rich MBE growth. Excess Al enhances the adatom surface migration process, leading to a 2D growth mode over a large area, while exhibiting parallel smooth atomic steps. These findings will enable the epitaxial growth of heterostructures for future electronic and photonic device applications based on the single-crystal N-polar AlN platform.

\begin{acknowledgements}
The authors at Cornell University acknowledge financial
support from Asahi Kasei, the Cornell Center for Materials Research (CCMR) — a NSF MRSEC program (No. DMR-1719875);
ULTRA, an Energy Frontier Research Center funded by the U.S.
Department of Energy (DOE), Office of Science, Basic Energy
Sciences (BES), under Award No. DE-SC0021230; AFOSR
Grant No. FA9550-20-1-0148; and Semiconductor Research Corporation (SRC) Joint University Microelectronics Program (JUMP). This work uses the Cornell Nanoscale Facilities, supported by NSF grant NNCI-202523 and CESI Shared
Facilities partly sponsored by NSF No. MRI DMR-1631282 and
Kavli Institute at Cornell (KIC).
\end{acknowledgements}

\section*{Data Availability Statement}

The data that support the findings of this study are available
from the corresponding author upon reasonable request.

\bibliography{N_polar_AlN}


\end{document}